\input epsf.def
\def\stacksymbols #1#2#3#4{\def\theguybelow{#2}
\def\verticalposition{\lower#3pt}
\def\spacingwithinsymbol{\baselineskip0pt\lineskip#4pt}
\mathrel{\mathpalette\intermediary#1}}
\def\intermediary#1#2{\verticalposition\vbox{\spacingwithinsymbol
\everycr={}\tabskip0pt
\halign{$\mathsurround0pt#1\hfil##\hfil$\crcr#2\crcr
\theguybelow\crcr}}}
\def\lapproxeq{\stacksymbols{<}{\sim}{2.5}{.2}}

\def\pmb#1{\setbox0=\hbox{#1}%
 \kern-.025em\copy0\kern-\wd0
\kern.05em\copy0\kern-\wd0
\kern-.025em\raise.0433em\box0 }

\newbox\figboxaa
\setbox\figboxaa\vbox{\hsize0.48\hsize
\epsfxsize3.5in
\centerline{\epsfbox{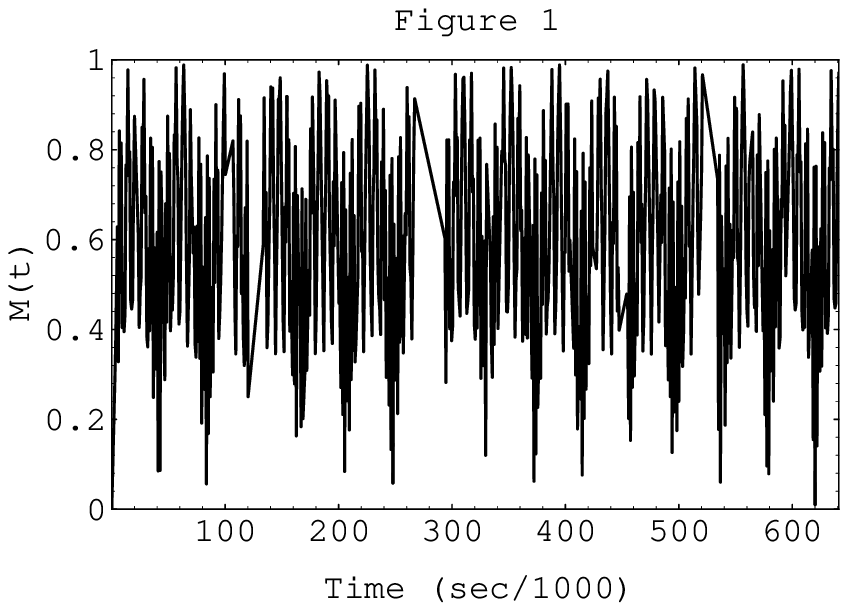}}
\vskip .1in
\noindent
{\bf FIG. 1.} Transverse magnetization 
for no entanglement ($j = 0$) and
no collapse ($\eta = 0$).
}

\newbox\figboxab
\setbox\figboxab\vbox{\hsize0.48\hsize
\epsfxsize3.5in
\centerline{
\epsfbox{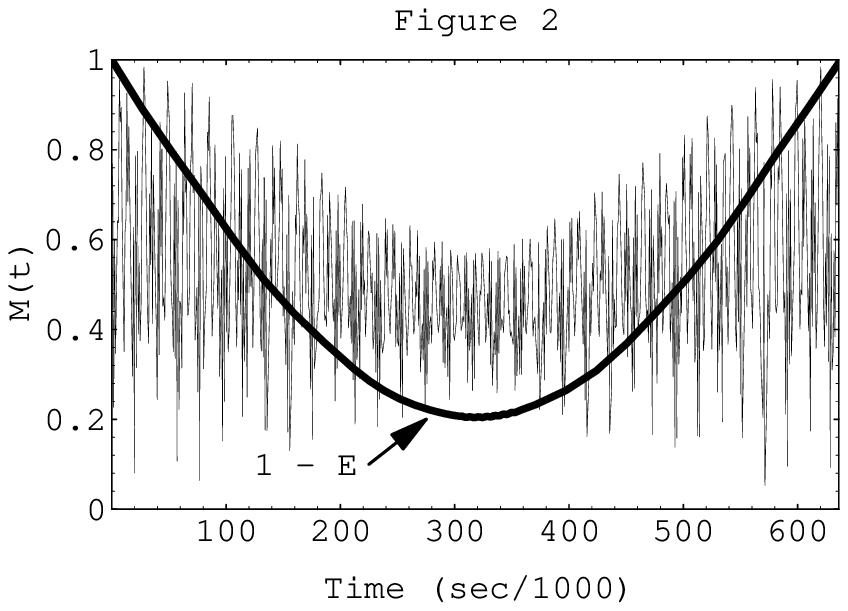}}
\vskip .1in
\noindent
{\bf FIG. 2.} Correlation between 
transverse magnetization and 
entanglement ($j = 0.0025$) without
collapse ($\eta = 0$).}

\newbox\figboxba
\setbox\figboxba\vbox{\hsize0.48\hsize
\epsfxsize3.5in
\centerline{
\epsfbox{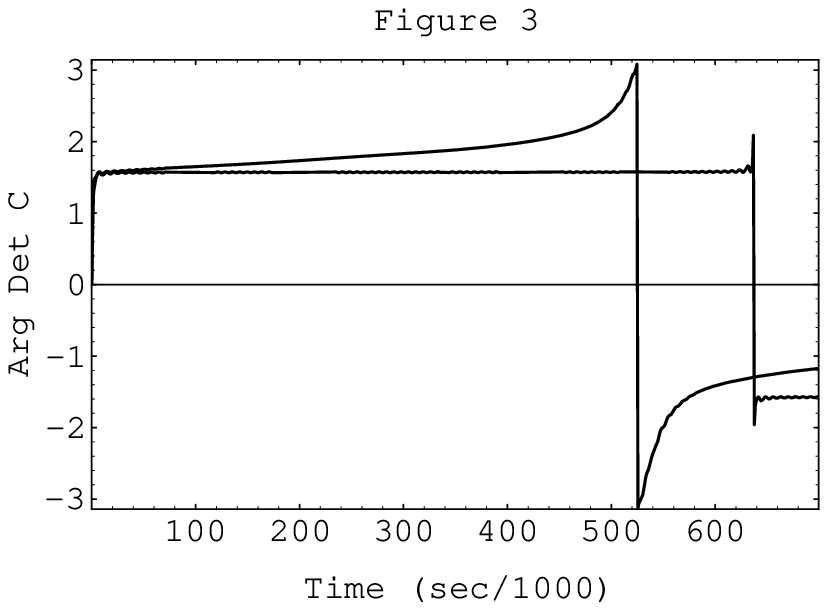}}
\vskip .1in
\noindent
{\bf FIG. 3.} Arg Det C for
$\eta = 0$ and $\eta = 2j$
}

\newbox\figboxbb
\setbox\figboxbb\vbox{\hsize0.48\hsize
\epsfxsize3.5in
\centerline{
\epsfbox{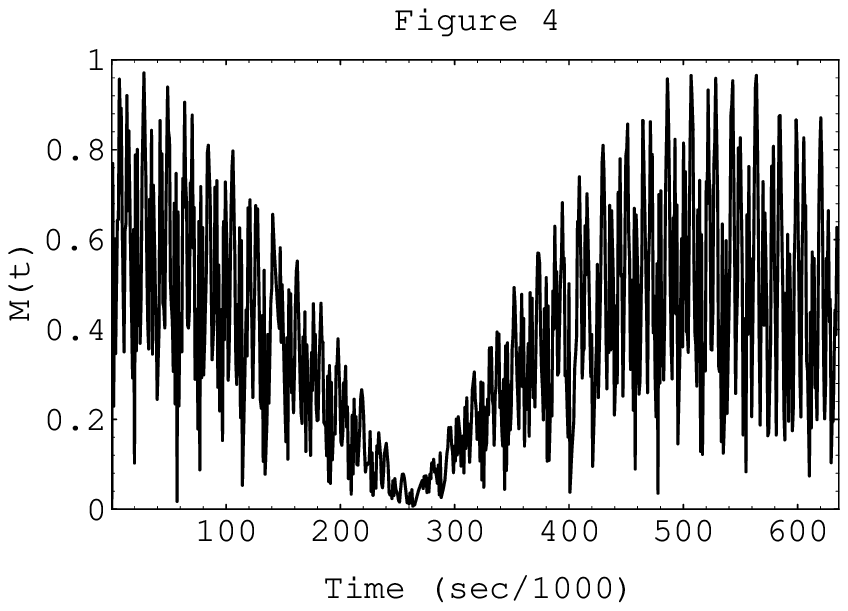}}
\vskip .1in
\noindent
{\bf FIG. 4.} Transverse magnetization 
when there is entanglement ($j = 0.0025$) and
INL collapse ($\eta = 2j$).
}

\centerline{{\bf How to Probe for Dynamical
Structure in the Collapse }}
{\centerline {{\bf of Entangled States
Using Nuclear Magnetic Resonance}}
\vskip.2in
\centerline{{\bf Daniel I. Fivel}}
\vskip.1in
\centerline{{\bf
Department of Physics, 
University of Maryland}}
\centerline{{\bf College Park, MD 20742}}
\centerline{{\bf May 29, 1997}}
\vskip.3in
\centerline{{\bf Abstract}}
\vskip.2in
The spin state of two magnetically inequivalent
protons in contiguous atoms of a molecule 
becomes entangeled by the indirect spin-spin
interaction (j-coupling). The degree of
entanglement oscillates at the beat frequency
resulting from the splitting of a degeneracy.
This beating is manifest in NMR spectroscopy  as
an envelope of the transverse magnetization and 
should be visible in the free induction decay
signal. The period ($\approx$ 1 sec) is long
enough for interference 
 between the linear dynamics and  collapse of the wave-function induced by a Stern-Gerlach inhomogeneity to
significantly alter the shape of that envelope. Various dynamical collapse theories can  be distinguished
by their observably different predictions with respect to this alteration. Adverse effects of detuning
due to the Stern-Gerlach inhomogeneity can be reduced to an acceptable level by having a sufficiently thin sample or
a strong rf field.
\vskip.1in
\noindent
Pacs:03.65.Bz - {\bf quant-ph 9706002}
\vskip.3in
The measurement problem of quantum mechanics$^{\bf 1}$ arises because the theory is
silent as to when and how the transition takes place in dynamical processes
between  linear, deterministic
Schr\"{o}dinger evolution, and  non-linear, stochastic wave-function collapse. One can sweep this
problem under the rug only so long as experimental probes are unable to enter the
transition region. Advancing techniques in mesoscopic physics$^{\bf 2}$ make such
an entry possible. The purpose of this paper is to suggest that we may already be
able to investigate the transition region using  well-developed techniques of
nuclear magnetic resonance (NMR). 
\vskip.1in
To see how we might go about distinguishing one theory of wave function collapse from another,
 let us suppose that we can arrange a dynamical transformation which, during a time $t_e$, causes
the following 
evolution of the spin state of two spin-1/2 particles:
$$ |\downarrow\rangle|\downarrow\rangle \to  2^{-1/2}(|\downarrow\rangle|\downarrow\rangle +
|\uparrow\rangle|\uparrow\rangle) \to |\downarrow\rangle|\downarrow\rangle.\eqno{(1)}$$
\vskip.1in
Suppose further that during this evolution an inhomogeneous
(Stern-Gerlach) magnetic field is used as a measuring device, the result of the measurent being that the system
collapses into  one of the states 
$|\downarrow\rangle|\downarrow\rangle$  or
$|\uparrow\rangle|\uparrow\rangle$. Spatial separation produced by
a Stern-Gerlach field is
proportional to the square of the time, and the spreading of the initially localized wave
function is linear in the time, so there is a time $t_{sg}$ at which the ratio
becomes unity, and we have a resolution of the two states.
\vskip.1in
Consider now what will happen if we arrange to have:
$$ t_e \approx t_{sg}.\eqno{(2)}$$
Two types of entanglement occur: the 
spin-spin entanglement occurring in (1) and  the entanglement of spin and space parts of the wave
function as the latter separates into non-overlapping fragments under the influence of
the Stern-Gerlach field.  One type of theory of wave-function collapse known as ``spontaneous localization theory"
(SLT)$^{\bf 3,4}$ attributes all collapse to spatial separation. In
such theories the entangled state
 $2^{-1/2}(|\downarrow\rangle|\downarrow\rangle +
|\uparrow\rangle|\uparrow\rangle)$ is not  driven to collapse 
 until the spatial separation has taken place. The situation is quite different, however, in another
 type of theory$^{\bf 5}$
known as ``induced non-linearity theory" (INL). In such theories any linear hamiltonian which is able to
distinguish the factorized constituents of an entangled state  must induce a non-linear term
that brings about collapse. If INL theory were correct, the presence of the inhomogenous magnetic field would
induce a non-linear term that would push the entangled spin-spin state towards collapse  {\it as that state is
forming}. Competition with the linear hamiltonian would thus occur all during the interval $0 < t < t_{sg}$.
Stopping  process (1) at various times
$t$ in this interval leaves the system in  different  states for the two theories, and this
difference  can have observable consequences.
\vskip.1in
We are now going to show that this scheme for testing a collapse theory experimentally can
be implemented using two proton spins in contiguous atoms $A$ and $A^\prime$ of a
molecule $AA^\prime$.
\vskip.1in
To begin with we must distinguish the sort of measurement we have in mind from conventional 
NMR spectroscopy$^{\bf 6,7}$. NMR spectroscopy probes the internal dynamics by
detecting shfts in the Larmor frequency
$\omega$ in a strong static magnetic field B. If  one applies a transverse magnetic field of magnitude $b << B$ 
in  a narrow frequency band including $\omega$, the dynamical effect is as if one moved to
a frame rotating with the Larmor frequency. In this frame the static field appears much weaker, and the transverse
field appears to be static. The internal dynamics then competes with the weakened magnetic interaction, and
the eigenfrequencies of the resulting hamiltonian carry information about the dynamics. Since these frequencies
also control the fourier distribution of the transverse magnetization, they can be detected by turning off the
transverse field and fourier analyzing the free induction decay signal which appears as the spins relax. It is
essential to maintain a high degree of homogeneity of the static field to prevent detuning of the Larmor 
frequency. But the testing of collapse dynamics {\it requires} that the field be inhomogeneous, and hence we 
cannot simply make use of small frequency shifts as a probe.
\vskip.1in
To obtain such a probe  we must find an experimental manifestation of entanglement that is
relatively insensitive to detuning. Fortunately, as we shall see, such a manifestation will be found in
a remarkable correlation between the degree of entanglement and a low frequency envelope of 
of the transverse magnetization (ETM). This correlation appears because both the entanglement
and the ETM are produced by the same mechanism:
 a {\it beating} between two frequencies split by  spin-spin interaction:
\vskip.1in
The dynamical situation of the two protons that we are going to exploit is the following:
A difference in their electronic environments  produces  a different ``chemical shift" of their
 magneto-gyric ratios and thereby of their
Larmor frequencies. This effect is proportional to the B field strength and typically $\sim$ 10ppm (parts per
million). There is a still smaller but measurable effect of the indirect spin-spin
interaction (j-coupling) which is transmitted from one spin to the other via a distortion in the electronic wave
function. The j-coupling, being a two-particle interaction, will entangle  a spin-spin state provided that
the state is not one of its eigenstates. Because of the chemical-shift the coupling to an external  magnetic
field is not simply proportional to a component of the angular momentum, and hence it does not commute with the
j-coupling. Thus the state is not an eigenstate of the j-coupling and will become entangled. The frequency shift
due to j-coupling is of order $j \approx 1 Hz$ (it is independent of the external field) and so the entanglement
time  $t_e$ will be of the order of seconds. 
\vskip.1in
Let us compare this with the Stern-Gerlach separation time $t_{sg}$.
For  $dB/dz \approx$ 1T/cm
we have
$$ t_{sg}^{-1} = { \gamma D \over \hbar} {dB\over dz},\eqno{(3)} $$ 
where $\gamma$ is the average magneto-gyric ratio for the two protons and $D$ is the molecular diameter.
 One finds that this is also of the
order of seconds, so  we have the required condition $t_{sg}\approx t_e$.
\vskip.1in
Our next step is to examine the relationship between the entanglement and the transverse magnetization
$$
M = (\langle\psi |\hbox{\pmb{$\sigma$}}_x /2|\psi\rangle^2 + 
\langle\psi|\hbox{\pmb{$\sigma$}}_y /2|\psi\rangle^2)^{1/2}.\eqno{(4)}.$$
 As shown in reference (5) there is a natural measure of entanglement:
$$E(\psi) \equiv  2 |\det C| = 2|C_{11}C_{22} - C_{12}C_{21}|,\eqno{(5)}$$
for the general state:
$$ |\psi(t)\rangle = 
C_{11}(t) |\uparrow\rangle |\uparrow\rangle +  C_{22}|\downarrow\rangle|\downarrow\rangle + C_{12}(t)
|\uparrow\rangle|\downarrow\rangle +  C_{21}(t) |\downarrow\rangle |\uparrow\rangle . \eqno{(6)}$$
One verifies that $E$ ranges from 0 for factorized states to 1 for maximally entangled states. 
To compute E and M we must next determine the evolution of the state $|\psi(t)\rangle$ with and without
the INL contribution.
\vskip.1in
 The hamiltonian describing two protons with magneto-gyric ratios $\gamma_1,\gamma_2$  in a
magnetic field ${\bf B}$ is
$$ {\cal H}_{\bf B} = ({\gamma_1\over 2}\hbox{\pmb{$\sigma$}}^{(1)}  + 
{\gamma_2\over 2}\hbox{\pmb{$\sigma$}}^{(2)})\cdot {\bf B} = 
{1 \over 4}((\gamma_1 + \gamma_2)\hbox{\pmb{$\sigma$}} + 
(\gamma_1 - \gamma_2)\delta \hbox{\pmb{$\sigma$}}))\cdot{\bf B}, $$
$$ \hbox{\pmb{$\sigma$}}\equiv \hbox{\pmb{$\sigma$}}^{(1)} + 
\hbox{\pmb{$\sigma$}}^{(2)},\quad  \delta \hbox{\pmb{$\sigma$}} \equiv \hbox{\pmb{$\sigma$}}^{(1)} -
\hbox{\pmb{$\sigma$}}^{(2)} .\eqno{(7)}$$ 
The term  with $\gamma_1 - \gamma_2$  in ${\cal H}_{\bf B}$ is the chemical
shift. We take the static contribution  to be of magnitude B in the z-direction and the rf contribution
to  be of magnitude $b$, rotating in the x-y plane with frequency $\omega$. The j-coupling hamiltonian
will be taken to be:
$$
{\cal H}_j = j\hbox{\pmb{$\sigma$}}^{(1)}\cdot\hbox{\pmb{$\sigma$}}^{(2)}.\eqno{(8)}$$
Going to a frame rotating with the frequency $\omega$ of the rf field, we obtain the effective
spin part of the hamiltonian:
$$
{\cal H}^\prime = {1\over 2}\{ \nu\hbox{\pmb{$\sigma$}}_z + d \, \delta \hbox{\pmb{$\sigma$}}_z + 2
j\,{\hbox{\pmb{$\sigma$}}}^{1}\cdot{
\hbox{\pmb{$\sigma$}}}^{2}
 +  \lambda\, \hbox{\pmb{$\sigma$}}_x \},$$
$$
\nu = \bar{\omega} - \omega,\qquad d = {1\over 2}(\omega_1 - \omega_2),\quad \lambda = (b/B)\bar{\omega}.
  $$
$$
\omega_1 = \gamma_1 B,\; \omega_2 = \gamma_2 B,\quad \bar{\omega} \equiv {1\over 2} (\omega_1 +
\omega_2). \eqno{(9)}$$
Here we have assumed that $b << B$ and so omitted the contribution of the chemical shift to the rf
term. 
The parameter $\nu$, called the ``detuning", indicates the difference  between the mean Larmor frequency
of the two protons and that of the rf oscillation.  Since we keep $\omega$ fixed, the variation in $\nu$
comes from the difference in the value of $\bar{\omega}$ for molecules in different parts of the sample
due to the inhomogeneity of the Stern-Gerlach field. 
\vskip.1in
The matrix
$$
{\cal H}^\prime =  \left(\matrix{ j  - \nu &  0 &  \lambda/2 & \lambda/2 \cr 
                                  0 &  j  + \nu  &  \lambda/2 & \lambda/2 \cr 
                                 \lambda/2 &  \lambda/2 & - j  + d &  2 j\cr
                                 \lambda/2 & \lambda/2 & 2 j &    - j - d}\right)
 \eqno{(10)}$$
represents ${\cal H}^\prime$ in the basis:
$$|1\rangle =
|\uparrow\rangle|\uparrow\rangle,\;|2\rangle = |\downarrow\rangle|\downarrow\rangle,\; |3\rangle =
|\uparrow\rangle|\downarrow\rangle,\; |4\rangle =|\downarrow\rangle|\uparrow\rangle. \eqno{(11)}$$ 
\vskip.1in
The period of $E$ can be deduced quite simply: For $j = 0$ the odd terms in the 
characteristic polynomial for ${\cal H}^\prime$ vanish, and so the eigenvalues have the form
 $\pm \kappa_0,\; \pm \kappa_1 $ with $0 < \kappa_0 < \kappa_1$. With $j > 0$ these are perturbed, and
in the parameter range above we find to first approximation:
$$
\pm \kappa_0 \Longrightarrow \pm \kappa_0 + j,
\qquad  \pm \kappa_1 \Longrightarrow \pm \kappa_1 - j . \eqno{(12)}$$
Now observe that when we compute $E$ we will encounter sums of products of two
exponentials of the form $e^{i\kappa t}$ and $e^{i \kappa^\prime t}$ where $\kappa,\kappa^\prime$ are any of the
four eigenvalues given in (12). One sees that there is only one low frequency that can be obtained from a
combination $\kappa + \kappa^\prime $, namely $\pm 2j$. In fact one obtains to lowest order in $j$:
$$E =  (1 + \nu^2/\lambda^2)^{-1} |\sin(2 j t)| \Longrightarrow t_e = \pi/(2j), \eqno{(13)}$$
  provided that $j << \nu$. Thus one
sees that the entanglement results from beating between the pair  $|\kappa_0 \pm j|$ or between the pair
$|\kappa_1 \pm j|$. If the initial state is $|\downarrow\rangle|\downarrow\rangle$ one finds that the
dominant contribution comes from the former.
\vskip.1in
These observations provide an important piece of information about the range of the detuning
parameter
$\nu$ in which we shall be interested. Because $\kappa_0 \to 0$ if
$\nu \to 0$ the beating is greatly reduced for zero  detuning. But the factor $(1 +
\nu^2/\lambda^2)^{-1}$ in (13) also reduces the entanglement as $\nu$ becomes comparable to $\lambda$, and so
there is a middle range between $j$ and $\lambda$ in which the entanglement is
 significant.
\vskip.1in
We next perform a numerical computation to examine the relationship between M and E for 
a choice of parameters that are expected to produce observable correlation. For protons at 
 B = 1T ($\bar{\omega} \approx $ 40 MHz) a chemical shift of 10 ppm corresponds to d = 400 Hz. It is convenient to
choose our energy or time unit such that $d = 1$. We choose an rf field with b = 1G which gives 
$\lambda = 10$ in our units, and we choose $j = 1Hz$ which is 0.0025 in our units. We choose $\nu = 5$
as typical for the range $1 \lapproxeq \nu \lapproxeq \lambda$.
\vskip.1in
In Figure 1 the transverse magnetization M is plotted in the absence of $j$-coupling so that there is no
entanglement. In Figure 2 with $j = 0.0025$  the disentanglement (1 - E) falls from unity to a minimum and then
rises back to unity in approximately 1.5 sec, and  {\it the envelope of the transverse magnetization is visibly
correlated with it}.
\vskip.2in
We next determine how the envelope is altered if,
as the INL theory asserts, there is competition
between the collapse mechanism and the linear
dynamics during the interval $0 < t < t_e$ when
$t_e \approx t_{sg}$. To do so we must first
apply the INL theory of reference (5) to deduce
the non-linear term appropriate to the experiment
under consideration: The strength
$\eta$ of the term is determined by the reciprocal of the time $t_{sg}$ required for the measuring device to
recognize the factorized constituents of the spin-spin entangled state. Having arranged that $t_{sg} \approx t_e$, it
follows from (13) that we must choose $\eta \approx 2j$.
Since the Stern-Gerlach  effect acts primarily on the states $|\uparrow\rangle|\uparrow\rangle,
|\downarrow\rangle|\downarrow\rangle$ we may let the  non-linear term act only in the subspace of the four
dimensional spin space  spanned by  these two states.  The stochastic element of
the theory is introduced by a random fluctuation in the sign of the non-linear term.  This  also occurs in times 
$ \sim t_{sg}$ and, since the proposed experiment is {\it confined} to $0 <t < t_{sg}$,  we may ignore this fluctuation.
\vskip.1in
The non-linear, stochastic Schr\"{o}dinger equation of reference (5) as applied to the experiment
under consideration can now be reduced to:
$$\hskip-.1in
{d\over dt}\left(\matrix{ C_{11}\cr C_{22}\cr C_{12} \cr C_{21} }\right) = - i
\left(\matrix{ j  - \nu &  0 &  \lambda/2 & \lambda/2 \cr 
                     0 &  j  + \nu  &  \lambda/2 & \lambda/2 \cr 
                   \lambda/2 &  \lambda/2 & - j  + 1 &  2 j\cr
              \lambda/2 & \lambda/2 & 2 j &    - j - 1}\right)\left(\matrix{ C_{11}\cr C_{22}\cr C_{12}
\cr C_{21} }\right)  + \eta \; e^{i\; arg\det C} \left(\matrix{0 &-1 & 0 & 0\cr 1 & 0 & 0 & 0 \cr
0 & 0 & 0 & 0 \cr 0 & 0 & 0 & 0\cr}\right)\left(\matrix{ C_{11}^*\cr C_{22}^*\cr C_{12}^* \cr C_{21}^* }\right).
\eqno{(15)}$$
\vskip.1in
\noindent
The factor $e^{i \; arg\det C}$ becomes indeterminate for $\det C \to 0$, i.e.\ when the state
factorizes. At this point the non-linear term turns off, which means that we may evaluate the expression as zero
when it becomes indeterminate.  The complex conjugation operation in
the last term is an essential feature of the INL theory, the significance of which will be discussed
below.
\vskip.1in
The task of integrating (15) is greatly simplified in our parameter range by the observation that
 $arg\det C$ is very nearly a constant during $0 < t < t_{sg}$. The reason is the following: 
Suppose first that $\eta = 0$. If we compute $C$ in the
basis  where ${\cal H}^\prime$ is diagonal, we find that $\arg\det C =  \pi/2$  
at all times. Since $j$ is
small, the transformation to that basis can be approximated by a tensor product of two one-particle
transformations, and such  transformations leave
$\det C$ invariant. Thus for $\eta = 0$ we conclude that $\arg\det C$  remains  close
to $\pi/2$ for $0 < t < t_e$. Since $\eta \approx 2j << 1$ in our computations, we expect that this
conclusion remains valid when the non-linear term is present. 
Suppose tentatively that this is the case. Then
the four equations (15)
together with the four complex conjugate equations are a {\it linear} system of eight equations
which can be readily solved. We can then check the validity of our approximation by 
self-consistency, i.e.\ computation of $arg\det C$ from the linearized solution and verification
that it is indeed approximately  $\pi/2$. 
\vskip.1in
 Figure 3 gives $arg\det C$ for $\eta = 0$ (the near step function), the other curve being a plot of
 $arg\det C$ calculated in the linearized theory for $\eta = 2j$. Evidently the approximation is self-consistent
 in the first half of the interval, though deviations 
from the predictions of the linearized equation may be expected in the second half.
\vskip.1in
Figure 4 shows the transverse magnetization for $\eta = 2j$ as calculated from the linearized
equation. Its salient feature is a strong depression of the envelope as time approaches midpassage.
This feature is not an artifact of the linear approximation because, as we have shown, that
approximation is accurate during the first half of the period.
\vskip.1in
 Thus we arrive at a principal 
result of this
paper: {\it INL theory predicts a descent of the envelope of the  transverse magnetization
 at times approaching the middle of the entanglement period.}
\vskip.1in
This descent will be found so long as  the detuning is in the range $1 \lapproxeq \nu \lapproxeq
\lambda$. If we assume that the gradient of the Stern-Gerlach field is $dB/dz = 1$ T/cm, we will
have such a range over 90\% of the sample if the thickness is 1 micron. This restriction can be
relieved by raising the strength of the rf field. Thus if b = 10G the sample size can be
increased to 10 microns with a corresponding increase in the effective signal. Ideally an experiment
should first verify the presence of an envelope in the transverse magnetization that has the
form of Figure 2 in  a homogeneous B field, and then ascertain whether or not it deforms to that of Figure 4 when
$dB/dz$ is of the order $1 T/cm$. The size of the deformation will place a bound on $\eta$, with $\eta << 2j$
indicating failure of the INL theory. The absence of any change from Figure 2 when B is made inhomogenous would
be consistent with spontaneous localization or with the commonly held view that collapse takes place
instantaneously when one of the two factorized constituents is detected.
 \vskip.1in
To conclude we shall comment on the physics underlying the envelope distortion produced by
the INL collapse mechanism: It was shown in reference 5 that the INL theory predicts a small CP
violation in the decay of neutral K mesons which, with a very coarse model, comes within 20\% of the
experimental value. Neutral kaons are entangled states of 
$d\bar{s}$ and $s\bar{d}$ quarks, and the non-linear term is induced by the
 semi-leptonic decay modes which can distinguish the factorized constituents. The source
of the CP symmetry-breaking is the complex conjugation operation in the INL theory. Let us now
show that the strong distortion of the transverse magnetization seen in Figure 4 
is also a consequence of the complex conjugation operation in (15):
\vskip.1in
 Let the matrix appearing in the  non-linear term be
denoted
$\Upsilon$. It is 
anti-symmetric and has non-vanishing elements only in the (12) and (21) position. When we transform to
the eigenbasis of ${\cal H}^\prime$ it will be seen that $\Upsilon \to \Upsilon^\prime$ in which
the latter is still anti-symmetric and has small off-diagonal elements except in the positions
(12) and (21.
In this basis the linear term produces oscillations $e^{-i\alpha t}$ with
$\alpha$ one of the four eigenvalues
$\kappa_0 - j,-\kappa_0 - j, \kappa_1 + j,$ and $-\kappa_1 + j$. Because of the form of
$\Upsilon^\prime$ just noted, the non-linear term produces a large coupling between the
first two only. Were it not for the complex conjugation  in the non-linear term, this would
have little effect because of the large frequency mismatch ($2\kappa_0$) bertween $\kappa_0 - j$ and
$-\kappa_0 - j$.
But, because of the complex conjugation, the coupling is between $\kappa_0 - j$ and
$-(-\kappa_0 - j)$ so that the mismatch is only $2j$. We thus have a near resonance between
the linear and non-linear term and  so obtain a large effect from a small perturbation that
persists over the long entanglement period.
\vskip.1in
The origin of the complex conjugation operation is the time-reversal operator, the occurrence
of which in the INL theory expresses the fact that while there is no arrow of time in
microscopic processes, that symmetry is broken whenever a measurement is registered. 
 Thus, if the INL theory is correct, it is this most basic
attribute of the measurement process that  manifests itself both in the CP asymmetry described in
reference (5) and
in the ETM depression descibed in this paper.
\vskip.3in
{\it Aacknowledgments:}
 The author thanks  S. Bhagat and A. Dragt for helpful
discussions. 
\vskip.3in
\centerline{{\bf References}}
\vskip.3in
1. D.\ Z. Albert, {\it Quantum Mechanics and Experience}, (Campbridge
University Press 1992).

2. C.\  Monroe, D.\ M.\  Meekhof, B.\  E.\  King, and D.\  J.\  Wineland,
Science, {\bf 272}, 1131 (1996).

3. G.\ C.\  Ghirardi, A.\  Rimini, and T.\  Weber, Phys. Rev. {\bf D34},
470 (1986).

4. P.\  Pearle, Phys. Rev. A {\bf 39}, 227 (1989).

5. D.\ I.\  Fivel, Phys. Rev. A  --- to be published - July 1997

6. C.\ P.\ Slichter, {\it Principles of Magnetic Resonance }, (Springer-Verlag,
Berlin, 1985)

7. F.\ A.\ Bovey, {Nuclear Magnetic Resonance Spectroscopy, 2nd Ed}, 

\hskip.25in(Academic Press, Inc,
New York, 1988)

\vskip.2in
\centerline{{\bf Figure Captions}}
\vskip.2in
\noindent
Figure 1: Transverse magnetization for no
entanglement (j = 0) and a homogeneous B field.
\vskip.1in
\noindent
Figure 2: Correlation between transverse
magnetization and entanglement (j = .0025).
B - field homogeneous. 
\vskip.1in
\noindent
Figure 3: Arg Det C for $\eta = 0$ (the
near step function) and for $\eta = 2j$.
\vskip.1in
\noindent
Figure 4: Transverse magnetization when
there is entanglement ($j = .0025$) and INL
collapse ($\eta = 2j$.)

\end